# Vibratory motion of vortices near the surface of type II superconductor under the AC magnetic field

L. G. Mamsurova, K. S. Pigalskiy
N. N. Semenov Institute of Chemical Physics RAS, 117977 Moscow, Russia

W. V. Pogosov
Moscow Institute of Physics and Technology, 141700 Dolgoprudnyi, Moscow Region, Russia

*Abstract*—The vibrational contribution $\mu_v(H)$ to the dynamic magnetic permeability is calculated allowing for the spatial variation of the order parameter in the vortex cores. The behavior of calculated dependence $\mu_v(H)$ is analyzed for the case when the vortices oscillate near their positions corresponding to the equilibrium vortex lattice. It is shown that taking into account the suppression of the order parameter in the vortex cores leads to appreciable difference in the behavior of $\mu_v(H)$ as compared to the similar dependence obtained in the London approximation. The main reason of this difference is a change of the effective potential well in which the vibrational motion of the vortices takes place.

## I. Introduction

The vibrational contribution to the dynamic magnetic permeability is caused by the oscillatory motion of vortices when static magnetic field $H$ is modulated by a weak *ac* magnetic field (of low frequency $\omega$ and small amplitude $h$). This contribution is given by:

$$\mu_v = \lim_{h \to 0}\left( \frac{1}{2\pi h} \int_0^{2\pi/\omega} \cos(\omega t) \frac{dB}{dt} dt \right), \qquad (1)$$

where $B(t)$ is the magnetic induction. The positions of vortices change during their vibrational motion under the *ac* field. Therefore, the magnetic induction changes and, according to (1), this leads to a nonzero value of $\mu_v$. It is important to note that since the Meissner current in a near-surface layer of thickness $\sim \lambda$ exceeds the critical current, the pinning can be neglected.

Our previous experimental studies [1] and [2] (carried out mainly on YBaCuO single crystals) and the results of other authors (see, e.g., [3] and [4]) have shown that $\mu_v(H)$ demonstrates hysteretic behavior in the magnetic field $H$ ($\mu_v(H)$ curve for increasing field is lower than that for decreasing field). The dependence $\mu_v(H)$ corresponding to different temperatures exhibits scaling behavior (Fig. 1). This behavior demonstrates that the effect under study is unrelated to the presence of the defects in the sample. Really, the pinning mechanism of hysteresis would lead to the irreversibility of the opposite sign from the experimental one. So, the only cause of this effect is the existence of hysteresis of $B(H)$ due to the surface barrier. Neglecting pinning processes one can treat the vibrational motion of near-surface vortices theoretically and obtain an explicit expression relating the quantity $\mu_v(H)$ with the value of $B$. This problem was solved in the London approximation in [1] and [2]. A comparison of the calculated $\mu_v(H)$ curves with the experimental ones showed that in YBaCuO samples (both single-crystal and polycrystalline) the surface barrier is strongly suppressed as compared to expected for an ideal surface [2]. The lower branch of the hysteresis of $\mu_v(H)$ is described well by the theoretical curve corresponding to equilibrium values of the induction, while the presence of the surface barrier has a more substantial effect on the exit of the vortices from the superconductor.

In the present study we concentrate our attention specifically on the fact that the experimental curve $\mu_v(H)$ related to the increasing field corresponds to the behavior of the equilibrium dependence $B(H)$ even in the vicinity of the lower critical field $H_{c1}$. As follows from the results of studies [5], the theoretical curve of $B(H)$, calculated allowing for the spatial variation of the order parameter in the vortex cores, differs from the similar curve obtained in the London approximation. In addition, the use of more accurate expression for the field of the vortex, taking into account the structure of its core, can lead to the change of $\mu_v(H)$ as compared to that calculated within the London model. In this context, it is of interest to obtain a relation for $\mu_v(H)$ taking into account the suppression of the order parameter in the vortex cores. In this paper we obtain such a relation and analyze the behavior of $\mu_v(H)$.

## II. Theory

Consider a superconducting sample in the form of a massive slab of thickness $D > \lambda$ in an external static magnetic field $H$ parallel to its surface. In accordance with [6], one can assume with high accuracy that the vortex rows are regularly spaced by a distance $d$ starting with the second one from the surface. The distance between the neighbor vortices within a row is denoted as $a$. We shall consider the field range $H \ll H_{c2}$, when the intervortex distances are much larger than the vortex core size. Then $\mu_v(H)$ is given by [2]:

$$\mu_v = \frac{2\Phi_0}{Da\lambda} \frac{\exp(-x_1/\lambda)}{1 - \exp(-d/\lambda)} \left(\frac{\partial x_1}{\partial H}\right)_{B_v}, \qquad (2)$$

where $x_1$ is the distance from the surface to the first vortex row, $B_v$ is the magnetic induction produced by vortices. It follows from (2) that $\mu_v(H)$ is determined by $x_1$. The equilibrium $x_1$ can be found by minimization of the Gibbs free energy $G$ per unit length of the vortex located in the first row. $G$ is given by:

This work is supported by the Russian Foundation for Basic Research (projects nos. 00-03-32246, 00-02-18032, and 01-02-06526) and by the Russian State Program 'Integracia'.

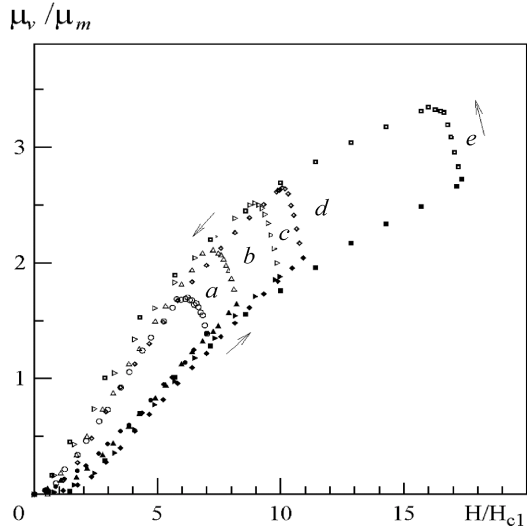

**Fig. 1.** Normalized experimental $\mu_v(H)$ curve for a YBa$_2$Cu$_3$O$_x$ single crystal at different temperatures $T$ [K]: 70 (*a*), 73.4 (*b*), 77.3 (*c*), 80 (*d*), 83.8 (*e*). The arrows indicate the directions of change of the magnetic field; $\mu_m$ is the normalization factor $\mu_m = 2\lambda/D$, where $D$ is the thickness of the sample.

$$G = F_0 + F_{em} - \frac{4\pi H}{\kappa}, \qquad (3)$$

where $F_0$ is the energy related to the variation of the order parameter in the vortex core, $F_{em}$ is the electromagnetic energy. In (3) and below we use the dimensionless variables: the distance $r$ and the magnetic field are measured in units of $\lambda$ and $H_c\sqrt{2}$, respectively, where $\lambda$ is the London penetration depth and $H_c$ is the thermodynamic critical field. In this notation we have: $H_{c2} = \kappa$, $\Phi_0 = 2\pi/\kappa$, where $\Phi_0$ is the magnetic flux quantum. Using the second Ginzburg-Landau equation, one can obtain: $F_{em} = \frac{2\pi}{\kappa}b(x_1,0)$, where $b(x_1,0)$ is the total magnetic field at the center of a vortex situated in the first row. In order to meet the boundary condition we use the method of images. The local field $b$ is written as the sum of the field $b_v$ of all the vortices, the field $b_i$ of their images, and the field $b_m$ of Meissner currents:

$$b = b_v + b_i + b_m, \qquad (4)$$

$$b_v = \sum_{vortices} b_0, \qquad b_v = \sum_{images} b_0, \qquad (5)$$

$$b_m = H\exp(-x), \qquad (6)$$

where $b_0$ is the local field of an isolated vortex, $x$ is the distance from the surface. To describe the local vortex field we use expression obtained in [7] by the variational method:

$$b_0(r) = \frac{K_0(\sqrt{r^2 + \xi_v^2})}{\kappa \xi_v K_1(\xi_v)}, \qquad (7)$$

where $K_n$ are the modified Bessel functions and $\xi_v = \sqrt{2}/\kappa$ is the parameter characterizing the vortex core size. Function (7) describes the results of an exact numerical solution of the Ginzburg-Landau equations with good accuracy. Since $F_{em}$ is the only term on the right-hand side of (3) depending on $x_1$, we obtain an equation for $x_1$:

$$\frac{\partial b(x_1,0)}{\partial x_1} = 0. \qquad (8)$$

The value of $b(x_1,0)$ is calculated by straightforward summation according to (4)-(7). The resulting equation for determining $x_1$ is:

$$\frac{B_v d}{2\xi_v K_1(\xi_v)}\left(\frac{\exp(-d)+\exp(-2x_1)}{1-\exp(-d)} - \frac{\xi_v^2}{2}(g(x_1)+f(d))\right) =$$
$$= H\exp(-x_1), \qquad (9)$$

where we introduced the notations:

$$g(x_1) = \frac{1}{4x_1^2} + \frac{1}{(2x_1+d/2)d}, \quad f(d) = \frac{5}{3d^2}.$$

Here for the simplicity we consider the square vortex lattice. We found numerically that the solution of (9) corresponding to equilibrium function $B_v(H)$ can be approximated by explicit expression:

$$x_1(H) = d/2 + \text{arcosh}\left(\frac{H}{\widetilde{B}_v}\right), \qquad (10)$$

where the following notation was used:

$$\widetilde{B}_v = \frac{B_v}{\xi_v K_1(\xi_v)}\left(1 - \frac{d^2}{24} - \frac{5\xi_v^2}{6d}\right). \qquad (11)$$

The calculation of $\mu_v(H)$ in the case of small amplitude of the *ac* field reduces to the problem of oscillations of the first vortex row near its equilibrium position [2]:

$$x_1(t,H) = x_1(H) + \frac{h\exp[-x_1(H)]}{2k\kappa}\sin(\omega t), \qquad (12)$$

where $k$ is the elastic constant, which governs the steepness of the effective potential well in which the oscillations of the vortices take place. Under the assumption that $k$ is determined mainly by the interaction of the vortices with each other and with the Meissner current and not by pinning ($k = \partial^2 G/\partial x_1^2$), we found:

$$k = \frac{1}{2\kappa}\exp(-d/2)\frac{\widetilde{B}_v\sqrt{H^2-\widetilde{B}_v^2}}{H+\sqrt{H^2-\widetilde{B}_v^2}}. \qquad (13)$$

Finally, we have:

$$\mu_v = \frac{2}{D}\frac{\widetilde{B}_v^2}{(H^2-\widetilde{B}_v^2)^{1/2}[H+(H^2-\widetilde{B}_v^2)^{1/2}]}. \qquad (14)$$

$\mu_v(H)$ depends on the term $H^2 - \widetilde{B}_v^2$. This term is much less than both $H(B_v)$ and $\widetilde{B}_v(H)$, therefore, some modifications of $H(B_v)$ and $\widetilde{B}_v(H)$ related to the vortex cores can lead to the appreciable change of $\mu_v(H)$, as compared to the London approach. Thus, (11) and (14) together with $H(B_v)$ give the dependence $\mu_v(H)$. Note that (9)-(14) remain valid in the London model if one puts $\xi_v = 0$.

Let us analyze the dependences $\mu_v(H)$ calculated within the developed approximation and within the London model. The equilibrium $B_v(H)$ allowing for vortex cores was found in [5] by means of the variational method. The resulting curve $\mu_v(H)$ calculated using this dependence and (14), (11) is plotted in Fig. 2 (curve 1). Within the London

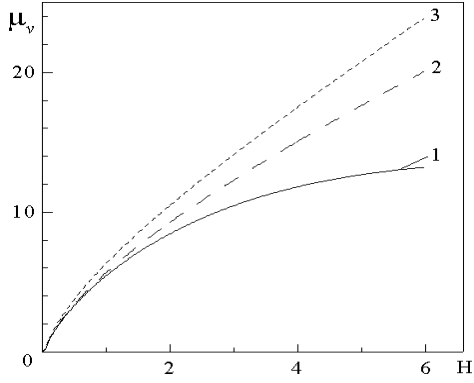

**Fig. 2.** Theoretical $\mu_v(H)$ dependences for the equilibrium vortex lattice at $\kappa = 100$. Curve 1 corresponds to the developed approximation where the vortex cores are taken into account. Curve 2 and 3 correspond to the London model with exact (16) and approximate (17) $H_{c1}$, respectively; $\mu_v$ and $H$ are measured in units of $2\lambda/D$ and $H_c\sqrt{2}$, respectively.

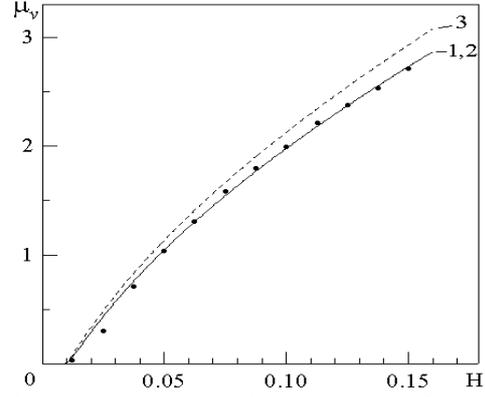

**Fig. 3.** The dependences $\mu_v(H)$ in the vicinity of $H_{c1}$. The points show experimental data, the numbers of the curves correspond to the numbers on the similar curves in Fig. 2; $\mu_v$ and $H$ are measured in units of $2\lambda/D$ and $H_c\sqrt{2}$, respectively.

model, the equilibrium relation between $B_v$ and $H$ was found in [8]:

$$H = B_v + H_{c1} - \frac{1}{4\kappa}\{\ln[2\kappa(H - H_{c1})] + 1.34\}. \quad (15)$$

The exact value of $H_{c1}$ was found in [9] by numerical solution of the Ginzburg–Landau equations:

$$H_{c1} = \frac{1}{2\kappa}(\ln\kappa + 0.5). \quad (16)$$

The resulting London curve $\mu_v(H)$ calculated using (14), (11) at $\xi_v = 0$ and (15), (16) is shown in Fig. 2 (curve 2). This curve differs from the results of the more accurate calculations of $\mu_v(H)$ allowing for the structure of the vortex cores (curve 1). The deviation of curves 1 and 2 is mainly due to the difference in the description of the local vortex field in the London model and in the developed model. The vortex field determines the energy of interaction of vortices with each other and with their images or the shape of the effective potential well, in which the vibrational motion of near-surface vortices takes place.

In order to illustrate the sensitivity of $\mu_v(H)$ to $B_v(H)$ and to the contributions from the vortex cores we calculate $\mu_v(H)$ within the London model (Fig. 2, curve 3) using also the approximate value of $H_{c1}$:

$$H_{c1} = \frac{K_0(1/\kappa)}{2\kappa}. \quad (17)$$

This value of $H_{c1}$ is often used and is found from the London expression for the local vortex field neglecting the contribution from the variation of the order parameter in the vortex core. It is seen from Fig. 2 that curves corresponding to different $H_{c1}$ values (curves 2 and 3) deviate starting from low fields. Note that curves 1 and 2 coincide in the vicinity of $H_{c1}$. This implies that in this field range taking the self-energy of the vortex into account by using the exact expression for $H_{c1}$ is enough for obtaining accurate $\mu_v(H)$ even in the London approximation. At higher fields the energy of interaction of vortices and images should be taken into account more accurately as compared to the London model.

Fig. 3 shows the calculated curves $\mu_v(H)$ in comparison with the experimental data measured under conditions of increasing magnetic field at $T = 84$ K for a YBaCuO single crystal in the geometry $H \perp c$. The technique used to obtain the experimental curves is described in detail in [1]. We see that the experimental data are in good agreement with the calculated curves 1 and 2. A criterion of good agreement is the value of the single fitting parameter: the London penetration depth. We obtained that $\lambda_{cp} = \sqrt{\lambda_{ab}\lambda_c} = 0.85$ mkm (which corresponds to $\lambda_{ab} = 0.38$ mkm) at $T = 84$ K. Although this value differs (by a factor of 1.4) from the known bulk values of $\lambda$ for YBaCuO single crystals [10], it is in good agreement with the surface values [11]. The similar results are obtained at other temperatures.

In summary, we showed that the effect of the vortex cores on the vibrational contribution $\mu_v$ to the dynamic magnetic permeability is essential even at low fields. In the vicinity of $H_{c1}$ it can be taken into account within the London model if one uses the exact $H_{c1}$, where the structure of the vortex core is already taken into account. At higher applied fields the energy of interaction of vortices with each other and with their images should be taken into account more accurately as compared to the London model. Using more adequate approach we calculated $\mu_v(H)$, accounting for the vortex cores.